\documentclass[aoas,MSNbibl,nameyear,dvips]{arximspdf}
\usepackage{mathbh,multirow,dcolumn}
\usepackage{graphicx}

\doi{10.1214/13-AOAS694} 
\volume{8}
\issue{1}
\pubyear{2014}
\firstpage{120}
\lastpage{147}

\makeatletter
\newcolumntype{d}[1]{D{.}{.}{#1}}
\newcommand{\E}{\mathrm{E}}
\renewcommand{\c}{c}

\makeatother

\begin{document}
\begin{frontmatter}

\title{Separable factor analysis with applications to~mortality
data\thanksref{T1}}
\runtitle{Separable factor analysis}

\begin{aug}
\author[A]{\fnms{Bailey K.} \snm{Fosdick}\corref{}\ead[label=e1]{bfosdick@samsi.info}}
\and
\author[B]{\fnms{Peter D.} \snm{Hoff}\ead[label=e2]{pdhoff@uw.edu}}
\runauthor{B. K. Fosdick and P. D. Hoff}
\affiliation{Statistical and Applied Mathematical Sciences
Institute\\  and University of Washington}
\address[A]{Statistical and Applied Mathematical\\
\quad Sciences Institute\\
P.O. Box 14006\\
Research Triangle Park, North Carolina 27709-4006\\
USA\\
\printead{e1}} 
\address[B]{Department of Statistics\\
University of Washington\\
P.O. Box 354322\\
Seattle, Washington 98195-4322\\USA\\
\printead{e2}}
\end{aug}
\thankstext{T1}{Supported in part by NICHD Grant R01HD067509.}

\received{\smonth{11} \syear{2012}}
\revised{\smonth{8} \syear{2013}}

%
\begin{abstract}
Human mortality data sets can be expressed as multiway data arrays, the
dimensions of which correspond to categories by which mortality rates
are reported, such as age, sex, country and year.
Regression models for such data typically assume an independent error
distribution or an error model that allows
for dependence along at most one or two dimensions of the data array.
However, failing to account for other dependencies
can lead to inefficient estimates of regression parameters, inaccurate
standard errors and poor predictions.
An alternative to assuming independent errors is to allow for dependence
along each dimension of the array using a separable covariance
model.
However, the number of parameters in this model increases rapidly with
the dimensions
of the array
and, for many arrays, maximum likelihood estimates of the covariance
parameters do not exist. In this paper, we propose a submodel of the
separable covariance
model that
estimates the covariance matrix for each dimension as having
factor analytic structure. This model can be viewed as an extension of
factor analysis to array-valued data, as it uses a factor model to
estimate the covariance along each dimension of the array.
We discuss properties of this model as they relate to ordinary factor
analysis, describe maximum likelihood and Bayesian estimation methods,
and provide a
likelihood ratio testing procedure for selecting the factor model ranks.
We apply this methodology to the analysis of
data from
the Human Mortality Database,
and show in a cross-validation
experiment
how it outperforms simpler methods.
Additionally, we use this model to impute
mortality rates for countries that have no mortality data
for several years. Unlike other approaches, our methodology is
able to estimate similarities between the mortality rates of countries,
time periods and sexes, and use
this information to assist with the imputations.
\end{abstract}

%

\end{frontmatter}

\section{Introduction}\label{secintro}
\begin{figure}
\includegraphics{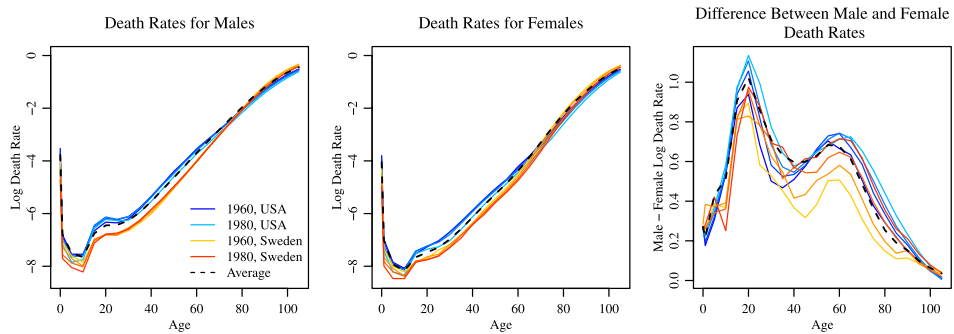}
\caption{Mortality curves for the United States of America and Sweden.
The gradient of colors for each country represents the log death rates
in the four 5-year time periods from 1960 to 1980. The average
sex-specific mortality curve over the four time periods and all
countries is shown in black.}\label{MortCurves}
\end{figure}

Human mortality data are used extensively by researchers and
policy makers to analyze historic and current population trends
and assess
long-term impacts of public policy initiatives.
To enable such inference,
numerous regression models have been proposed
that estimate mortality rates as a function of age using a small number
of parameters
[\citet{HP}, \citet{ModeBusby}, \citet{Siler}].
Practitioners using these methods typically
model the age-specific death rates for each country, year and sex
combination separately and assume independent error distributions.
Examples of death rates analyzed by such methods are shown in
Figure~\ref{MortCurves} for the United States and Sweden. Each
mortality curve is defined by 23 age-specific death rates and the
average sex-specific mortality curve from 1960--1980 over thirty-eight
countries is also displayed.

From the figure, it is clear that a country's mortality rates in one
time period
are similar to its rates in adjacent time periods.
Acknowledging this fact,
several researchers have developed models for
``dynamic life tables,'' that is, matrices of mortality rates for
combinations of ages and time periods,
for single country--sex combinations. An example of such a life table is
the male death rates in Sweden from 1960 to 1980 shown in Figure~\ref
{MortCurves}.
Some of the models developed for these data specify ARIMA processes for
the time-varying model parameters [\citet{McNown1989},
\citet{RenshawHaberman2003age}],
while others
smooth the death rates over age and time using a kernel smoother
[\citet{Felipe2001}], $p$-splines [\citet{Currie2004}],
nonseparable age--time
period covariance functions [\citet{MartinezRuiz2010}] or multiplicative
effects for age and time [\citet{LeeCarter}, \citet
{Renshaw1996}, \citeauthor{RenshawHaberman2003par} (\citeyear{RenshawHaberman2003par,RenshawHaberman2003red}), \citet{ChiouMuller2009}].

%

Human mortality data sets typically provide mortality rates of
populations corresponding to combinations of several
factors. For example, the Human Mortality Database (HMD) [University of
California, Berkeley and Max Planck Institute for Demographic Research
(\citeyear{HMD})] provides mortality rates of populations corresponding to
combinations of 40 countries, 9 time periods, 23 age groups, and both
male and female sexes.
As is shown in Figure~\ref{MortCurves},
mortality rates of men and women within a country will typically both
be higher than or both lower than the sex-specific rates averaged
across countries. Furthermore,
differences between male and female mortality rates generally show
trends that are consistent across
countries and time periods.
Such patterns suggest
joint estimation of mortality rates
using
a model
that can
share information across
levels of two or more factors.
Two models that consider death rates for more than one country or sex
are that developed by \citet{LiLee2005}, which estimates common
age and
time period effects for a group of countries or both sexes, and
\citet{CarterLee1992}, where male and female death rates within
the same
country share a time-varying mortality level. Although these methods
consider either both sexes or multiple countries,
the extreme similarity of the curves in Figure~\ref{MortCurves}
for males across countries and for a given country across sexes
suggest that separately modeling
death rates for different countries or sexes is inefficient, and
inference may be improved by using a~joint model that shares information across all factors.

\begin{figure}[t]
\includegraphics{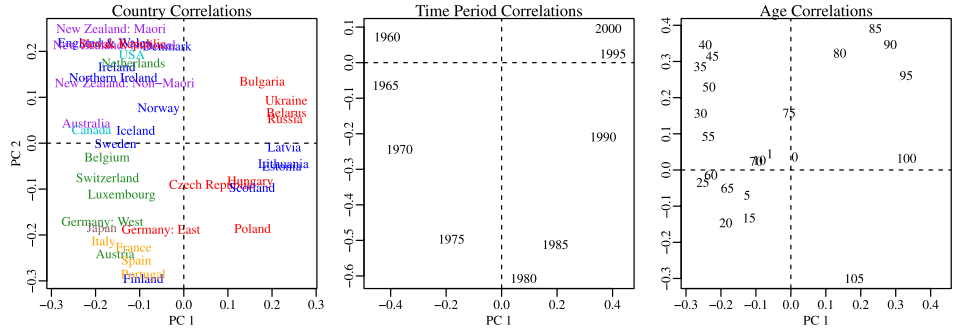}
\caption{The first two principal components of each sample correlation
matrix are displayed, and countries in the same United Nations region
are shown in the same color. Close proximity in the principal
components space away from the origin is indicative of a positive correlation.}\label{pcs}
\end{figure}

With this in mind,
we consider a regression model for the HMD data
consisting of a mean model that is a piecewise-polynomial in age with
additive effects for country, time period and sex
(more details on this model, and its comparison to other models,
are provided in Section~\ref{sechmd}).
This mean model
is extremely flexible:
it contains over 370 parameters and
an ordinary least squares (OLS) fit accounts for over 99\% of the total
variation in the data (coefficient of determination, $R^{2} > 0.995$).
Nonetheless,
an analysis of the residuals from the OLS fit indicates that
some clear patterns in the data are not captured by the regression model
and, in particular,
a model of independent errors is a poor
representation of these data.
To illustrate this, note that the residuals
can be represented as a
4-way array, the dimensions of which are given by the number of levels of
each of the four factors: country, time period, sex and age.
To examine residual correlation across levels of a factor, the
4-way array of residuals can be converted into a matrix
whose columns represent the levels of the factor, and
a sample correlation matrix for the factor can be obtained.
Figure~\ref{pcs} summarizes the patterns in the residual correlations
using the first two principal components of each sample correlation matrix.
If a model of independent errors were to be adequate,
we would expect the sample correlation values to be small and centered
about zero, and no discernible patterns to exist in the principal components.
However,
the sample
correlations are substantially more positive than would be expected
under independence: 59\% of the observed country correlations, 61\% of
time period correlations and 98\% of age correlations are greater than
the corresponding 95\% theoretical percentiles under the independence
assumption. Additionally, there are clear geographic, temporal and age
trends in the principal components in Figure~\ref{pcs}. For example,
the residuals for the Ukrainian
mortality rates are positively correlated to those for Russia, and
the residuals for the year 2000 are positively correlated with those for
1995. This residual analysis suggests that an assumption of
uncorrelated errors is inappropriate.

%

Failure to recognize correlated errors can
lead to a variety of inferential problems, such as
inefficient parameter estimates and inaccurate standard errors.
For the analysis of the mortality data, an additional important
consequence is that
the accuracy of predictions of missing mortality rates may suffer.
Predicting missing death rates is a primary application of modeling
mortality data, as developing countries often lack reliable death
registration data. It is possible that
the residual dependence could be reduced by increasing the flexibility of
the mean model,
but since this is already fairly complex,
we may instead prefer to
represent residual dependence with a covariance model, leading to
a general linear model for the data in which the
mean function and residual covariance are estimated
simultaneously.

The mortality
data, like the residuals, can be represented as a 4-way array, each dimension
of which corresponds to one of the factors of country, time period,
sex and age. In the literature on multiway array data
[see, e.g., \citet{kroonenberg2008}],
each dimension is referred to as a mode of the array, so the
4-way array of mortality data consists of four modes.
As described by \citet{HoffSep}, a natural covariance model for
a $K$-way data array
is
a separable covariance model, parameterized in terms of
$K$ covariance matrices, one for each mode of the array.
If the array is also assumed to be normally distributed, the model is
referred to as the array normal model
and can be seen as an extension of the matrix normal model [\citet
{Dawid}].

Even though the separable covariance model is not a full, unstructured
covariance model,
the array normal likelihood is unbounded
for many array dimensions, prohibiting the use of maximum likelihood
methods [\citet{Manceur2012}]. Estimates of the array normal covariance
parameters can still be obtained by taking a Bayesian approach
[\citet{HoffSep}] or
by using a penalized likelihood [\citet{GAllen}]. However, the
lack of
existence of the maximum likelihood estimates (MLEs) indicates that the
data is unable to provide information about all of the parameters. In
this article we propose an alternative modeling approach that
parameterizes the covariance matrix of each mode by a reduced rank
matrix plus a diagonal matrix, referred to here as factor analytic
covariance structure. This new model, called Separable Factor Analysis
(SFA), is an extension of factor analysis to array-valued data and
provides a parsimonious representation of mode-specific covariance
in an array-valued
data set. The reduction in the number of parameters by using covariance
matrices with factor analytic structure leads to existence of MLEs for
the SFA parameters in many cases when the MLEs of the array normal
parameters do not exist.

This article is outlined as follows: in the next section we introduce
and motivate SFA, as well as discuss its properties and similarities to
ordinary factor analysis. We describe two estimation procedures in
Section~\ref{secest}: an iterative maximum likelihood algorithm and a
Metropolis--Hastings sampler for inference in a Bayesian framework. A
likelihood ratio testing procedure for selecting the rank of the factor
model for each mode is also presented. In Section~\ref{sechmd} the SFA
model is used to analyze the HMD mortality data and its performance is
compared to simpler covariance models in a simulation study.
We
illustrate how SFA uses estimated similarities between country
mortality rates to
provide imputations for countries missing mortality data for
several years.
This prediction method extends the approach taken in \citet{CD},
\citet{Brass}, \citet{UN} and \citet{Murray}, where
one country's mortality
curve is modeled a function of another's.
Our approach is novel in that it estimates the covariance between
mortality rates across all countries, time periods and sexes, and uses
these relationships to impute missing death rates.
We conclude with a discussion in Section~\ref{secdis}.

\section{Extending factor analysis to arrays}\label{secSFA}

\subsection{Motivating separable factor analysis}
Suppose $Y$ is a $K$-way array of dimension $m_{1} \times m_{2} \times
\cdots\times m_{K}$. We are interested in relating the data $Y$ to
explanatory variables $X$ through the model $Y=M(X,\beta) + E$, where
$\beta$ represents unknown regression coefficients and $E$ represents
the deviations from the mean. As was discussed in the preliminary
analysis of the mortality data,
it is often unreasonable to assume the elements of $E$ are independent
and identically distributed.

In cases where there is no independent replication, estimation of the
$\operatorname{Cov}[E]$ can be problematic, as it must be based on
essentially a single sample. One solution is to approximate the
covariance matrix with one with simplified structure. A frequently used
model in spatio-temporal analysis is a separable covariance model
[\citet{Stein}, \citet{Genton}], which estimates a
covariance matrix for each
mode of the array. It is written $\operatorname{Cov}[\operatorname
{vec}(E)] = \Sigma_{K} \otimes\Sigma_{K-1} \otimes\cdots\otimes
\Sigma
_{1}$, where ``vec'' and ``$\otimes$'' denote the vectorization and
Kronecker operators, respectively. In the context of the mortality
data, this model contains a covariance matrix for country ($\Sigma
_{\c}$), time period ($\Sigma_{t}$), age ($\Sigma_{a}$) and sex
($\Sigma
_{s}$). A separable \mbox{covariance} model with the assumption that the
deviations are normally distributed, $\operatorname{vec}(E) \sim
\operatorname{normal}(0,\operatorname{Cov}[\operatorname{vec}(E)])$, is
an array normal model and was developed by \citet{HoffSep} as an
extension of the matrix normal [\citet{Dawid}, \citet
{Browne}, \citet{Oort}].

The mode covariance matrices in the array normal model are not
estimable for certain array dimensions using standard techniques such
as maximum likelihood estimation [\citet{Manceur2012}]. However, often
the covariance matrices of large modes can be well approximated by
matrices with simpler structure. A common approach in the social
sciences to modeling the covariance of a high-dimensional random vector
$x \in\mathbb{R}^{p}$ is to use a $k$-factor model, which
parameterizes the covariance matrix as $\operatorname{Cov}[x] =
\Lambda
\Lambda^{T} + D^{2}$, where $\Lambda\in\mathbb{R}^{p \times k}$, $k <
p$, and $D$ is a diagonal matrix [\citet{Spearman}, \citet
{MKB}]. We will
refer to this model as single mode factor analysis, as it models the
covariance among one set of variables. When the number of independent
observations $n$ is less than~$p$, the sample covariance matrix is not
positive definite and hence cannot be used as an estimate of
$\operatorname{Cov}[x]$. Nevertheless, under the assumption that $x$
follows a multivariate normal distribution with known mean, the maximum
likelihood estimate of the factor analytic covariance matrix exists if
$k < \min(p,n)$ [\citet{RobertSymons}].

We propose a submodel of the array normal model where each mode
covariance matrix potentially has factor analytic structure. We call
this model \textit{Separable Factor Analysis} (SFA) and it is written
as follows:
%
\begin{eqnarray}
\operatorname{vec}(E) &\sim& \operatorname{normal} \bigl
(0,\operatorname{Cov}
\bigl[\operatorname{vec}(E) \bigr] \bigr),
\nonumber
\\
\operatorname{Cov} \bigl[\operatorname{vec}(E) \bigr] &=& \Sigma_{K}
\otimes\Sigma_{K-1} \otimes\cdots\otimes\Sigma_{1},
\label{SFA}
\\
\mbox{where } \Sigma_{i} &=& \Lambda_{i}\Lambda_{i}^{T}
+ D_{i}^{2}\qquad\mbox{for } 0 \le k_{i} <
m_{i}
\nonumber
\end{eqnarray}
and $\Sigma_{i}$ is unconstrained (i.e., equals any positive definite
matrix) if $k_{i}=m_{i}$. SFA models are characterized by the
covariance matrix structure chosen for each mode and can be represented
by a $K$-vector of ranks $(k_{1},\ldots, k_{K})$, where $k_{i}$ equals
the rank of $\Lambda_{i}$ if mode $i$'s covariance matrix has factor
analytic structure and equals $m_{i}$ if the mode covariance matrix is
unstructured. Note that we consider the $k_{i}=0$ case where the
covariance matrix is diagonal. A key advantage of the SFA model over
the array normal model is that empirical evidence has shown that MLEs
of the SFA covariance parameters exist for array dimensions where the
MLEs of the array normal unstructured covariance matrices do not exist.

\subsection{Properties of SFA}\label{secmode}
In this section we relate the SFA parameters to those in ordinary
factor analysis, discuss indeterminancies in the model, and interpret
the SFA parameters when the true covariance matrix in each mode is
unstructured. This requires the concept of array matricization. Here we
follow the convention set in \citet{Kolda} where the
matricization of an
array in the $i$th mode is defined as the ($m_{i} \times\prod_{j \neq
i} m_{j})$ matrix $Y_{(i)}$, whose column indices vary faster for
earlier mode indices than later mode indices [see \citet{Kiers}
and \citet{HOSVD} for alternative definitions].
\begin{longlist}
\item[\textit{Latent variable representation}.]
Although the primary motivation for the factor analytic structure of
the mode covariance matrices in SFA is parameter reduction, the SFA has
a convenient latent variable formulation similar to that in single mode
factor analysis. A single mode $k$-factor model for a sample of $n$
mean-zero $p$-variate random vectors is written $\{x_{1},\ldots,x_{n}\}
\sim\mbox{i.i.d. } \operatorname{normal}(0,\Lambda\Lambda^{T} +
D^{2})$, where $\Lambda\in\mathbb{R}^{p \times k}$ and $D$ is a
diagonal matrix. Defining $X=[x_{1},\ldots,x_{n}]$ as the $p \times n$
matrix of observations, this model has an equivalent latent variable
representation as a decomposition into common latent factors,
$Z=[z_{1},\ldots,z_{n}]$, and variable specific latent factors,
$E=[e_{1},\ldots,e_{n}]$, as follows:
%
\begin{eqnarray}\label{m1}
X_{p \times n} &=& \Lambda_{p \times k} Z_{k \times n} + D_{p \times p}
E_{p \times n},
\nonumber
\\
\{z_{1},\ldots,z_{n} \} &\sim&
\mbox{i.i.d. }\operatorname{normal}(0,\mathrm{I}_{k}),\qquad
\operatorname{Cov}[z_{i},e_{j}]=0_{k \times p}
\nonumber\\[-10pt]\\[-10pt]
\eqntext{\mbox{for all $i,j$},}
\\
\{e_{1},\ldots,e_{n} \} &\sim&\mbox{i.i.d. }\operatorname
{normal}(0,\mathrm{I}_{p}).
\nonumber
\end{eqnarray}
This representation expresses the $j$th observation of the $i$th
variable $X_{ij}$ as a linear combination of common latent factors
$z_{j}$ with coefficients given by the $i$th row of~$\Lambda$, plus a
single variable specific factor $E_{ij}$, scaled by the $i$th diagonal
element of $D$.

A similar representation exists for each mode with a factor analytic
covariance structure in the SFA model. Consider a mean-zero array $Y$
and an SFA model with a factor analytic covariance matrix in the $i$th
mode. Define $\widetilde{Y}{}^{i}$ to be the array obtained by
standardizing $Y$ with all but the $i$th mode's covariance matrix:
%
\begin{equation}
\qquad\operatorname{vec} \bigl(\widetilde{Y}{}^{i} \bigr):=
\operatorname{vec}(Y) \bigl(\Sigma_{K}^{-1/2} \otimes\cdots\otimes
\Sigma_{i+1}^{-1/2} \otimes\mathrm{I}_{m_{i}} \otimes
\Sigma_{i-1}^{-1/2} \otimes\cdots\otimes\Sigma_{1}^{-1/2}
\bigr). \label{removeOneFA}
\end{equation}
It follows that
%
\begin{eqnarray}\label{FAm2LV}
\{y_{1},\ldots,y_{m_{-i}} \} &\sim&\mbox{i.i.d. }\operatorname
{normal} \bigl(0,
\Lambda_{i}\Lambda_{i}^{T} + D_{i}^{2}
\bigr)\quad\mbox{and}
\nonumber\\[-8pt]\\[-8pt]
\widetilde{Y}{}^{i}_{(i)}&=&[y_{1},\ldots,y_{m_{-i}}] \buildrel d
\over= \Lambda_{i} Z^{i}
+ D_{i} E^{i},\nonumber
\end{eqnarray}
where $m_{-i}= \prod_{j \neq i} m_{j}$, and $Z^{i}$ and $E^{i}$ are
$k_{i} \times m_{-i}$ and $m_{i} \times m_{-i}$, respectively, with the
same distributional properties as $Z$ and $E$ in (\ref{m1}). The
superscript $i$ on $\widetilde{Y}{}^{i}_{(i)}$ indicates the $i$th mode
has not been standardized and the subscript $(i)$ indicates the array
has been matricized along the $i$th mode. The representation in $(\ref
{FAm2LV})$ suggests the parameters $\{\Lambda_{i}, D_{i}\}$ can be
viewed as single mode factor analysis parameters for the $i$th mode of
the array when the covariance in all other modes has been removed. This
representation is used in the parameter estimation methods in
Section~\ref{secest}.
\end{longlist}
\begin{longlist}
\item[\textit{Model indeterminacies}.]
SFA as parameterized in (\ref{SFA}) has two indeterminacies, one of
which is common to all factor models and one that is common to all
array normal models. The first indeterminacy, which is also present in
single mode factor analysis, is the orientation of the $\Lambda$
matrices. The array covariance matrix in (\ref{SFA}) is the same with
mode $i$ factor analytic parameters $\{\Lambda_{i}, D_{i}\}$ as it is
with parameters $\{\Lambda_{i}G_{i}, D_{i}\}$, where $G_{i}$ is any
$k_{i} \times k_{i}$ orthogonal matrix. A common identifiable
parameterization of $\Lambda$ is that which restricts $\Lambda$ to be
lower-triangular with positive diagonal elements [\citet
{GewekeZhou1996}, \citet{Carvalho2008}; see \citet
{AndersonRubin} for
alternative identifiability conditions]. The formulation in (\ref{SFA})
can be viewed as a model with parameter-expanded $\Lambda_{i}$
matrices, similar to that in \citet{BhattacharyaDunson}, since it
includes no identifiability constraints.

The second indeterminacy concerns the scales of the mode covariance
matrices and stems from the model's separable covariance structure. For
example, the transformation $\{\Sigma_{i},\Sigma_{j}\} \mapsto\{
c\Sigma
_{i},\Sigma_{j}/c\}$ does not affect the array covariance matrix in~(\ref{SFA}) for any $c>0$. This scale nonidentifiability is eliminated
if all mode covariance matrices are restricted to have trace equal to
one and a scale parameter is included for the total variance of the array.
\end{longlist}
\begin{longlist}
\item[\textit{Pseudo-true parameters}.]
In single mode factor analysis the goal is to represent the covariance
among a large set of variables in terms of a small number of latent
factors. However, often it is unlikely the true covariance matrix
$\Sigma$ has factor analytic structure. Therefore, there is interest in
what $k$-factor analytic parameter values, $\Lambda$ and $D$, best
approximate the true covariance matrix $\Sigma$. These optimal
parameter values, denoted $\overline{\Lambda}(\Sigma)$ and $\overline{D}(\Sigma)$, are those that minimize the Kullback--Leibler (KL)
divergence between the $k$-factor model and the multivariate normal
model. Minimizing the KL divergence is equivalent to maximizing the
expected value of the $k$-factor analysis (FA) probability density with
respect to the true multivariate normal (MN) distribution. Letting
$X=[x_{1},\ldots,x_{n}]$ where $\{x_{1},\ldots,x_{n}\} \sim\mbox
{i.i.d. }\operatorname{normal}(0,\Sigma)$, $\overline{\Lambda
}(\Sigma)$
and $\overline{D}(\Sigma)$ can be defined as
\begin{eqnarray*}
\bigl\{\overline{\Lambda}(\Sigma), \overline{D}(\Sigma) \bigr\}
&:=&\mathop{\arg\max}\limits
_{\Lambda, D} \E_{\mathrm{MN}} \bigl[p_{\mathrm
{FA}}(X|\Lambda,D)
\bigr]
\\
&=& \mathop{\arg\max}\limits_{\Lambda, D} c_{\mathrm{FA}}-\frac
{n}{2} \log\bigl(\bigl|\Lambda
\Lambda^{T} + D^{2}\bigr| \bigr) - \frac{n}{2}
\operatorname{tr} \bigl[ \bigl(\Lambda\Lambda^{T} + D^{2}
\bigr)^{-1}\Sigma\bigr],
\end{eqnarray*}
where ``tr'' represents the trace operator and $c_{\mathrm{FA}}$ is a
constant not\vspace*{1pt} depending on~$\Lambda$~or~$D$. In the case of $k=0$, the
best approximating diagonal matrix $\overline{D}{}^{2}$ contains the
diagonal elements of $\Sigma$.

Similarly, SFA is an approximation to a separable covariance structure
where modes' true covariance matrices are unlikely to have factor
analytic structure. Suppose the distribution\vspace*{2pt} of $Y$ is array normal
with mean zero and covariance matrices $\widetilde{\bolds{\Sigma}}
= \{\widetilde{\Sigma}_{i}\dvtx 1 \le i \le K \}$. Consider a
$(k_{1},\ldots,k_{K})$ SFA model for $Y$ with parameters $\bolds
{\Lambda} = \{
\Lambda_{i}\dvtx 0 < k_{i} < m_{i}\}$, $\mathbf{D}=\{D_{i}\dvtx 0
\le k_{i}
< m_{i}\}$ and $\bolds{\Sigma}= \{\Sigma_{j}\dvtx k_{j} = m_{j}\}$.
The expected value of the SFA probability density with respect to the
true array normal (AN) model is
%
\begin{eqnarray}\label{exploglike}
\E_{\mathrm{AN}} \bigl[ p_{\mathrm{SFA}}(Y|\bolds{\Sigma},
\mathbf{D},\bolds{\Lambda}) \bigr] &=& c_{\mathrm{SFA}} - \sum
_{i=1}^{K}\frac{m}{2m_{i}} \log\bigl(|
\Sigma_{i}|\bigr) - \frac{1}{2} \prod_{i=1}^{K}
\operatorname{tr} \bigl[\Sigma_{i}^{-1}\widetilde{\Sigma}_{i}
\bigr]
\nonumber\\[-8pt]\\[-8pt]
\eqntext{\mbox{where } \Sigma_{i} = \Lambda_{i}\Lambda_{i}^{T}
+ D_{i}^{2}\qquad\mbox{for } 0 \le k_{i} <
m_{i},}
\end{eqnarray}
$c_{\mathrm{SFA}}$ is a constant independent of the SFA parameters, and
$m = \prod_{i=1}^{K} m_{i}$. Let $\overline{\overline{\Lambda
}}(\widetilde{\bolds{\Sigma}})$, $\,\overline{\overline{\!D}}(\widetilde{\bolds{\Sigma}} )$ and $\overline{\overline{\Sigma
}}(\widetilde{\bolds{\Sigma}} )$ denote the SFA parameters that
maximize (\ref{exploglike}) and, hence, provide the best approximation
to the true separable covariance matrix based on $\widetilde
{\bolds{\Sigma}}$. It can be shown that for all appropriate $i$, $j$
and $k$,
%
\begin{equation}
\overline{\overline{\Lambda}}_{i}(\widetilde{\bolds{\Sigma}}) =
\overline{\Lambda}(\widetilde{\Sigma}_{i}),\qquad\,\overline{\overline{\!D}}_{j}(\widetilde{\bolds{\Sigma}}) = \overline{D}(
\widetilde{\Sigma}_{j})\quad\mbox{and}\quad\overline{\overline{
\Sigma}}_{k}(\widetilde{\bolds{\Sigma}} ) = \widetilde{
\Sigma}_{k}. \label{optSFA}
\end{equation}
This implies that the best factor analytic parameters, $\{\overline{\overline{\Lambda}}_{i}(\widetilde{\bolds{\Sigma}} ), \,\overline{\overline{\!D}}_{j}(\widetilde{\bolds{\Sigma}})\}$, for a given
mode in the SFA model are the closest\vspace*{1pt} fitting single mode factor
analytic parameters to that mode's true covariance matrix, $\{{\overline{\Lambda}}_{i}({\widetilde{\Sigma}}_{i} ), {\overline{D}}_{j}({\widetilde{\Sigma}}_{i})\}$. As we might expect, the optimal\vspace*{1pt}
values of the unstructured covariance matrices in the SFA model, $
\overline{\overline{\Sigma}}_{k}(\widetilde{\bolds{\Sigma}} )$,
are the modes' true covariance matrices $\widetilde{\Sigma}_{k}$.

This implies that when the true model is array normal, the optimal SFA
parameters for a given mode do not depend on the specified covariance
structures in the other modes. Note that the scale indeterminacy of the
covariance matrices is still present here, such that there is a set of
optimal SFA parameter values that provide the same approximation.
Asymptotically, as the number of replicates of the array increases,
these optimal SFA parameter values are the limiting values of the SFA
maximum likelihood estimates [\citet{White1982}].
\end{longlist}

\section{Estimation and testing}\label{secest}
In this section we consider parameter estimation for the SFA model and
propose a likelihood ratio testing procedure for selecting the ranks
$(k_{1},\ldots,k_{K})$. Two estimation methods are described here: an
iterative algorithm for maximum likelihood estimation and a
Metropolis--Hastings algorithm which approximates the posterior
distribution of the parameters given the data. We present the case
where the array has mean zero, however, both estimation methods and the
testing procedure can be extended to allow for simultaneous estimation
of a mean structure and the SFA covariance structure. Examples of such
extensions are discussed in Section~\ref{sechmd} for the mortality data.

\subsection{Maximum likelihood estimation} \label{secMLE}
While simultaneous maximization of the SFA log likelihood with respect
to all parameters is difficult, maximizing the log likelihood with
respect to a single mode's covariance parameters is feasible. Thus, we
propose a block coordinate ascent algorithm that iteratively maximizes
the SFA log likelihood over a single mode's covariance parameters using
the latest values of all other modes' parameters and is guaranteed to
increase the log likelihood at each step.

Let $\bolds{\Lambda} = \{\Lambda_{i}\dvtx 0 < k_{i} < m_{i}\}$,
$\mathbf{D}=\{D_{i}\dvtx 0 \le k_{i} < m_{i}\}$, and $\bolds{\Sigma
} = \{\Sigma_{j}\dvtx k_{j} = m_{j}\}$ as in Section~\ref{secmode}. Also,
let $\bolds{\Lambda}_{-j} = \bolds{\Lambda}/\{ \Lambda_{j} \}
$ be the set $\bolds{\Lambda}$ with $\Lambda_{j}$ removed, and
define $\bolds{D}_{-j}$ and $\bolds{\Sigma}_{-i}$
analogously. The iterative maximum likelihood algorithm proceeds as follows:
\begin{longlist}[4.]
\item[0.] Specify initial values for all covariance parameters $\{\bolds{\Lambda
},\mathbf{D},\bolds{\Sigma}\}$.
\item[1.]
For each mode $\{i\dvtx k_{i}=0 \}$, update the estimate of $D_{i}$.
\item[2.]
For each mode $\{i\dvtx 0 < k_{i} < m_{i} \}$, update the estimates of
$\Lambda_{i}$ and $D_{i}$.
\item[3.]
For each mode $\{i\dvtx k_{i}=m_{i} \}$, update the estimate of
$\Sigma_{i}$.
\item[4.]
Repeat steps 1--3 until a desired level of convergence is obtained.
\end{longlist}

The maximization of the SFA log likelihood for the updates in steps 1~and~3 are straightforward.
Differentiating the log likelihood with respect to $D_{i}$ or $\Sigma
_{i}$, it can be shown the updates for steps 1~and~3, respectively, are
\[
D_{i}^{2} = \operatorname{diag} \biggl( \frac{m_{i}}{m}
\widetilde{Y}{}^{i}_{(i)} \bigl(\widetilde{Y}{}^{i}_{(i)}
\bigr)^{T} \biggr)\quad\mbox{and}\quad\Sigma_{i} =
\frac{m_{i}}{m}\widetilde{Y}{}^{i}_{(i)} \bigl(\widetilde
{Y}{}^{i}_{(i)} \bigr)^{T},
\]
where the covariance matrices used to standardize $Y$ in $\widetilde
{Y}{}^{i}$ are the latest covariance matrix estimates and $m=\prod
_{i}^{K} m_{i}$.

Estimation of a mode's factor analytic parameters in step~2 is more
difficult, but can be accomplished using methods developed for single
mode factor analysis. The SFA log likelihood as a function of the $i$th
mode's factor analytic parameters~is
%
\begin{eqnarray}\label{faMLarr}
\ell(\Lambda_{i},D_{i}
|\bolds{\Sigma},\bolds{\Lambda}_{-i},\mathbf{D}_{-i},Y)
&=& c_{i} - \frac{m}{2m_{i}}\log\bigl( \bigl|\Lambda_{i}
\Lambda_{i}^{T}+D_{i}^{2}\bigr| \bigr)
\nonumber\\[-8pt]\\[-8pt]
&&{}-\frac{1}{2} \operatorname{tr} \bigl[ \bigl(\Lambda_{i}\Lambda
_{i}^{T}+D_{i}^{2} \bigr)^{-1}
\widetilde{Y}{}^{i}_{(i)} \bigl(\widetilde{Y}{}^{i}_{(i)}
\bigr)^{T} \bigr],\nonumber
\end{eqnarray}
where $c_{i}$ is a constant not depending on $\Lambda_{i}$ or $D_{i}$.
The log likelihood for a single mode $k_{i}$-factor model for a $p
\times n$ matrix $X$ is written
%
\begin{equation}
\ell(\Lambda,D |X) = c - \frac{n}{2}\log\bigl( \bigl|\Lambda\Lambda
^{T}+D^{2}\bigr| \bigr)-\frac{1}{2} \operatorname{tr} \bigl[
\bigl(\Lambda\Lambda^{T}+D^{2} \bigr)^{-1}
XX^{T} \bigr]. \label{faML}
\end{equation}
Notice that the SFA log likelihood has the same form as that for
single mode factor analysis where $XX^{T}$ and $n$ are replaced by
$\widetilde{Y}{}^{i}_{(i)} (\widetilde{Y}{}^{i}_{(i)})^{T}$ and $m/m_{i}$,\vspace*{1pt}
respectively. Therefore, estimation methods for single mode factor
analysis can be used to update $\Lambda_{i}$ and $D_{i}$ in step~2.

Numerous iterative algorithms have been developed to obtain the single
mode factor model maximum likelihood estimates, however, many suffer
from poor convergence behavior [\citet{Lawley}, \citet
{Joreskog}, \citet{NR}]. An expectation--maximization (EM)
algorithm was developed based
on the model representation in (\ref{m1}) that treats $Z$ as latent
variables [\citet{Dempster}, \citet{EM}]. The slow
convergence of this
algorithm led to expectation/conditional maximization either (ECME)
algorithms, some of which rely on numerical optimization procedures
[\citet{ECME}, \citet{ZYJ}]. \citet{ZYJ} proposed an
iterative algorithm
that updates $\Lambda$, treating $D$ as known, and then sequentially
updates each diagonal element of $D$, treating $\Lambda$ and all other
elements of $D$ as known. This algorithm has closed form expressions
for all parameter updates and was shown to outperform the EM algorithm
and its extensions in terms of convergence and computation time. For
these reasons, we chose to use it for step~2 of the SFA estimation procedure.

Divergence of the SFA maximum likelihood algorithm, where the log
likelihood continually grows at a nondecreasing rate, is evidence that
the maximum likelihood estimates do not exist. While the update in step~1 for a mode with a diagonal covariance matrix is always well defined
(i.e., the SFA log likelihood has a maximum in terms of $D_{i}$), step~2 of the algorithm for an unstructured covariance matrix is only well
defined if $m_{i}<\prod_{j \neq i} m_{j}$. Similarly, step~3 is well
defined for a mode $i$ if $k_{i}<\operatorname{rank} (\widetilde
{Y}{}^{i}_{(i)} (\widetilde{Y}{}^{i}_{(i)})^{T} )$. This latter
requirement is effectively equivalent to $k_{i} < \min(m_{i},\prod_{j
\neq i} m_{j})$ since $\widetilde{Y}{}^{i}_{(i)}$ is unlikely to be rank
deficient for a~continuous array $Y$.

Since no identifiability constraints are placed on the mode covariance
matrix scales or the factor analytic $\Lambda_{i}$ parameters, the
estimates that result from the above procedure correspond to a set of
equivalent estimates obtained by reallocating the scale of the
covariance matrices and rotating the $\Lambda_{i}$ matrices. If
interpretation of the $\Lambda$ matrices is of interest, an
identifiable parameterization can be obtained from the resulting
estimate using the restrictions mentioned in Section~\ref{secmode}. The
iterative maximum likelihood estimation procedure can be extended to
simultaneously estimate parameters $\beta$ associated with an array
mean model $M(X,\beta)$, if an additional step is added to the
procedure that maximizes the normal log likelihood with respect to
$\beta$ and $\widetilde{Y}$ is redefined as the array that has been
standardized by both the mean and covariance matrices.

\subsection{Bayesian estimation}\label{secMCMC}
Maximum likelihood estimates of the SFA covariance parameters and any
mean model parameters $\beta$ can be obtained using the block
coordinate ascent algorithm, however, obtaining standard errors of the
estimates based on the Fisher information matrix requires
complicated derivatives and large matrix inversion. While numerical
estimation of the information matrix is possible [\citet
{spall2005monte}], an alternative estimation procedure that readily
provides parameter uncertainty estimates is that based on a Bayesian
approach. In this framework inference for the parameters is based on
the joint posterior distribution of the parameters given the data,
$p(\bolds{\Lambda}, \mathbf{D}, \bolds{\Sigma}|Y) \propto
p(Y|\bolds{\Lambda}, \mathbf{D}, \bolds{\Sigma})p(\bolds{\Lambda
}, \mathbf{D}, \bolds{\Sigma})$, where
$p(Y|\bolds{\Lambda}, \mathbf{D}, \bolds{\Sigma})$ is the
density of the $(k_{1},\ldots,k_{K})$ SFA model and $p(\bolds{\Lambda
}, \mathbf{D}, \bolds{\Sigma})$ is the joint prior
distribution of the parameters. Again, we present this algorithm for
the mean-zero array case, however, it can be trivially extended to
include mean model parameters $\beta$.
\begin{longlist}
\item[\textit{Prior specification}.]
In the absence of real prior information, we suggest a convenience
prior composed of semiconjugate distributions for the parameters. For
each mode $i$ with an unstructured covariance matrix, the prior
distribution for $\Sigma_{i}^{-1}$ is $ \operatorname{Wishart}(\kappa
_{i}, \mathrm{I}_{m_{i}})$ with hyperparameter $\kappa_{i}$, where
$\kappa_{i} \ge m_{i}$. For a mode $i$ with a factor analytic
covariance matrix, the joint prior distribution of $\{\Lambda
_{i},D_{i}\}$ is specified as follows:
%
\begin{eqnarray}
\bigl\{ \operatorname{vec}(\Lambda_{i})|D_{i} \bigr\} &\sim&
\operatorname{normal} \bigl(0,\mathrm{I}_{k_{i}} \otimes
D_{i}^{2} \bigr), \label{Lprior}
\\
\qquad \bigl\{ D^{-2}_{i}[1,1],\ldots,D^{-2}_{i}[m_{i},m_{i}] \bigr\}
&\sim&\mbox{i.i.d. }\operatorname{gamma} \bigl(\nu_{0}/2,
\mbox{ rate}= \nu_{0}d_{0}^2/2 \bigr),
\label{Dprior}
\end{eqnarray}
where $\nu_{0}> 0$ and $d_{0}^{2} > 0$. A priori each mode's parameters
are modeled as independent of all other modes' parameters given the
hyperparameters $\nu_{0}$, $d_{0}^2$ and $\{ \kappa_{i}\dvtx
k_{i}=m_{i}\}$.

The prior distribution of the factor analytic parameters given in (\ref{Lprior})--(\ref{Dprior}) has nice properties related to the rotational
indeterminacies in the $\bolds{\Lambda}$ matrices. Recall that the
SFA likelihood is invariant to rotation of $\Lambda_{i}$, meaning
L$_{\mathrm{SFA}}(\Lambda_{i}, D_{i}, \bolds{\Sigma}, \bolds
{\Lambda}_{-i}, \mathbf{D}_{-i}|Y)=\mathrm{L}_{\mathrm
{SFA}}(\Lambda
_{i}G_{i}, D_{i}, \bolds{\Sigma}, \bolds{\Lambda}_{-i},
\mathbf{D}_{-i}|Y)$, where L$_{\mathrm{SFA}}$ is the SFA likelihood
and $G_{i}$ is any $k_{i} \times k_{i}$ orthogonal matrix. Integrating
the joint prior distribution $p(\Lambda_{i},D_{i})$ over $D_{i}$, the
marginal distribution of $\Lambda_{i}$ is obtained and can be
expressed as
%
\[
p(\Lambda_{i}) \propto\prod_{j=1}^{m_{i}}
\bigl[ \nu_{0}d_{0}^2 + \bigl\|\Lambda_{i}[j,]\bigr\|^{2}
\bigr] ^{(k_{i} + \nu_{0})/2},
\]
where $\|\cdot\|^{2}$ denotes the Frobenius norm. Observe that
$p(\Lambda_{i})=p(\Lambda_{i}G_{i})$, implying that the prior
distribution is also invariant to rotations of $\Lambda_{i}$. This is a
desirable property, as it indicates the prior does not favor one set of
parameters over another if they are equivalent given the data (i.e.,
have the same SFA likelihood).
\end{longlist}
\begin{longlist}
\item[\textit{Metropolis--Hastings algorithm}.]
The posterior distribution $p(\bolds{\Lambda},\mathbf{D},
\bolds{\Sigma} |Y)$ is not a standard distribution and is
difficult to sample from directly, so we propose approximating it using
samples from a Metropolis--Hastings algorithm. This algorithm produces a
Markov chain in $\{\bolds{\Lambda},\mathbf{D},\bolds{\Sigma}\}$,
whose stationary distribution is equal to $p(\bolds{\Lambda},\mathbf
{D}, \bolds{\Sigma}|Y)$, and proceeds by
iteratively proposing new values of each mode's parameters. Typically,
in such an algorithm, proposals are accepted based on a probability
that is a function of the data likelihood, prior and proposals,
however, the parameter proposals in this algorithm all have acceptance
probability equal to one. The algorithm can be described as follows:
\begin{longlist}[4.]
\item[0.]
Specify initial values for all covariance parameters $\{\bolds{\Lambda
},\mathbf{D},\bolds{\Sigma}\}$.
\item[1.]
For each mode $\{i\dvtx k_{i}=0 \}$, sample $D_{i}$ from its full
conditional distribution:
%
\begin{eqnarray}\label{FCd}
&& \bigl\{ D_{i}^{-2}[j,j]
| \bolds{D}_{-i},\bolds{\Lambda},\bolds{\Sigma},Y \bigr
\}
\nonumber\\[-8pt]\\[-8pt]
&&\qquad \sim\operatorname{gamma} \bigl( (\nu_{0}+m/m_{i})/2,
\mbox{ rate} = \bigl(\nu_{0}d_{0}^2 +
S_{i}[j,j] \bigr)/2 \bigr)\nonumber
\end{eqnarray}
for $j \in\{1,\ldots,m_{i} \}$ where $S_{i} = \widetilde
{Y}{}^{i}_{(i)}(\widetilde{Y}{}^{i}_{(i)})^{T}$.
\item[2.]
For each mode $\{i\dvtx 0 < k_{i} < m_{i} \}$, sample new values of
$\Lambda_{i}$ and $D_{i}$ using the following steps:
\begin{enumerate}[(a)]
\item[(a)]
Sample $\{\operatorname{vec}(Z^{i}) | \Lambda_{i}, D_{i},\widetilde
{Y}{}^{i}\} \sim\operatorname{normal}(\operatorname{vec}(\phi
\Lambda
_{i}^{T}D_{i}^{-2}\widetilde{Y}{}^{i}_{(i)}),\mathrm{I}_{m/m_{i}}
\otimes
\phi)$
where $\phi= (\Lambda_{i}^{T}D_{i}^{-2} \Lambda_{i}+\mathrm{I})^{-1}$.
\item[(b)]
Sample
\begin{eqnarray*}
\hspace*{-6pt}&& \bigl\{\operatorname{vec}(\Lambda_{i}) |Z^{i},Y,
\mathbf{D},\bolds{\Lambda}_{-i},\bolds{\Sigma} \bigr\}
\\
\hspace*{-6pt}&&\!\qquad \sim \operatorname{normal} \bigl(\gamma\bigl(Z^{i}_{(i)}
\otimes D_{i}^{-2} \bigr)\operatorname{vec} \bigl(\widetilde
{Y}{}^{i}_{(i)} \bigr),
\gamma = \bigl[ \bigl( Z^{i}_{(i)}
\bigl(Z^{i}_{(i)} \bigr)^{T} + \mathrm
{I}_{m_{i}} \bigr) \otimes D_{i}^{-2}
\bigr]^{-1} \bigr).
\end{eqnarray*}
\item[(c)]
Sample $\{\operatorname{vec}(Z^{i}) | \Lambda_{i}, D_{i},\widetilde
{Y}{}^{i}\}$ as in 2(a).
\item[(d)]
Sample the elements of $D^{2}_{i}$ independently from
\begin{eqnarray*}
\hspace*{-6pt}&& \bigl\{ D_{i}^{-2}[j,j] |Z^{i},Y,
\mathbf{D}_{-i},\bolds{\Lambda},\bolds{\Sigma} \bigr\}
\\
\hspace*{-6pt}&&\!\qquad \sim \operatorname{gamma} \bigl( (\nu_{0}+m/m_{i} +
k_{i})/2, \mbox{ rate} = \bigl(\nu_{0}d_{0}^2 +
J[j,j] + \bigl\|\Lambda_{i}[j,]\bigr\|^{2} \bigr)/2 \bigr),
\end{eqnarray*}
where $J = (\widetilde{Y}{}^{i}_{(i)} - \Lambda_{i}Z^{i})(\widetilde
{Y}{}^{i}_{(i)} - \Lambda_{i}Z^{i})^{T}$ and $\|\cdot\|^{2}$ denotes the
Frobenius norm.
\end{enumerate}
\item[3.]
For each mode $\{i\dvtx k_{i}=m_{i} \}$, sample $\Sigma_{i}$ from its full
conditional distribution:
%
\begin{equation}
\bigl\{ \Sigma_{i}^{-1}|\mathbf{D},\bolds{\Lambda},
\bolds{\Sigma}_{-i},Y \bigr\} \sim\operatorname{Wishart} \bigl(
\kappa_{i} + m/m_{i}, \bigl(\mathrm{I}_{m_{i}} +
\widetilde{Y}{}^{i}_{(i)} \bigl(\widetilde{Y}{}^{i}_{(i)}
\bigr)^{T} \bigr)^{-1} \bigr). \label{FCsig}
\end{equation}
\item[4.]
Repeat steps 1--3 until a sufficiently accurate approximation of the
posterior distribution is obtained.
\end{longlist}
The covariance matrices used to standardize $Y$ in $\widetilde{Y}{}^{i}$
in each of the updates above are the most current parameter updates.
The updates of the factor analytic parameters $\{\Lambda_{i}, D_{i}\}$
in step~2 are\vspace*{-1pt} based on the latent variable representation of SFA
introduced in (\ref{FAm2LV}), which\vspace*{-2pt} expresses $Y$ as $ \widetilde
{Y}{}^{i}_{(i)}\buildrel d \over= \Lambda_{i} Z^{i} + D_{i} E^{i} $,
where the elements of $Z^{i}$ and $E^{i}$ are independent standard
normal random variables. Proof that the acceptance probabilities are
equal to one for the proposals of $\Lambda_{i}$ and $D_{i}$ is provided
in \hyperref[appMH]{Appendix}. Note that the $Z^{i}$ variables involved in
step~2 are not included as parameters, as is done in a
parameter-augmented sampler, but instead are simply used to propose new
factor analytic parameters.

Since no identifiability restrictions are placed on the scales of the
mode covariance matrices or on the orientation of the $\Lambda$
matrices, the model can be viewed as a parameter-expanded model,
similar to the single-mode factor model in \citet{BhattacharyaDunson}.
Working with this parameterization greatly simplifies the estimation
procedure and avoids the index order dependence issues that arise from
performing estimation with an identifiable parameterization where
choice of the index order within a mode becomes an important modeling
decision [see \citet{Carvalho2008} and \citet
{BhattacharyaDunson} for
further discussion]. Note that identifiability of the $\Lambda$
matrices is irrelevant when the goal in the analysis is covariance
matrix estimation, mean model inference or prediction of missing
values. However, if interpretation of the factor analytic parameters is
of interest, the samples from the Markov chain can be transformed to
identifiable parameters using the restrictions mentioned in
Section~\ref{secmode}. Similarly, posterior inference on the total
variance of the
array can be based on the combined scales of each sample of covariance
matrices, obtained by scaling all mode covariance matrices to have
trace one.

Unlike in the frequentist setting where divergence of the maximum
likelihood estimation procedure indicates a lack of information in the
data about the parameters, the posterior distribution of the parameters
given the data will always exist. Although Bayesian parameter estimates
are available, we should be aware of what information the estimates
reflect. Extreme similarity between the prior distribution and the
posterior distribution suggests that little information is gained from
the data and inference based on the posterior distribution is primarily
a reflection of the information in the prior.
\end{longlist}
\begin{longlist}
\item[\textit{Hyperparameters}.]
When there is little prior information about the parameters, it is
common to choose hyperparameter values that result in diffuse prior
distributions. We propose $\nu_{0}=3$ and $\kappa_{i}=m_{i}+2$ for $\{
i\dvtx k_{i}=m_{i}\}$ as default values, as they correspond to prior
distributions whose first moments are finite and represent some of the
most diffuse distributions in the Wishart and gamma families,
respectively. They also have the following properties:
%
\begin{equation}
\E[\Sigma_{i}] = \mathrm{I}_{m_{i}},\qquad
\E \bigl[D_{i}^{2}[j,j] \bigr] = 3
d_{0}^2,\qquad \E \bigl[\operatorname{tr}
\bigl(\Lambda_{i} \Lambda_{i}^{T} \bigr) \bigr] = 3
k_{i}m_{i} d_{0}^2 \label{priorexp}.
\end{equation}

Prior information about specific mode covariance matrices may be
limited, however, an estimate $\hat{\psi}$ of the total variance of
the array, $\psi=\operatorname{tr}(\operatorname{Cov}[\operatorname
{vec}(Y)]) = \prod_{i=1}^{K} \operatorname{tr}(\Sigma_{i})$, may be
available. This information can improve parameter estimation by
centering the prior distribution of the total variance of the array
around a~reasonable value. Based on the expectations in (\ref
{priorexp}) and the independence of the mode covariance matrices in the
prior, the prior expected value of the total variance of the array will
equal the estimate, $\E[\operatorname{tr}(\operatorname
{Cov}[\operatorname{vec}(Y)])] = \hat{\psi} $, if
%
\begin{equation}
d_{0}^{2} =\hat{\psi} {}^{1/R} \Biggl[ \biggl(
\prod_{j\dvtx 0 < k_{j} <
m_{j}} [k_{j} + 1] \biggr) \Biggl(
\prod_{i=1}^{K} m_{i} \Biggr)
3^{R} \Biggr]^{-1/R}, \label{d2spec}
\end{equation}
where $R = \sum_{i=1}^{K} \mathbh{1}\{ 0 \le k_{i}<m_{i} \} $ is the
number of modes with factor analytic covariance structure. In the event
there is no prior knowledge about $\psi$ and it is not of interest in
the analysis, we propose taking an empirical Bayes approach and
obtaining an estimate of it based on the data.\vspace*{1pt} Possible estimates
include $\hat{\psi} = \|Y\|^{2}$ or $\hat{\psi} = \|Y-\widehat
{M}(X,\beta)\|^{2}$ if the model has a nonzero mean. In the latter
case, $\widehat{M}(X,\beta)$ represents an initial estimate of the
mean, such as the ordinary least squares estimate. A similar approach
was suggested in \citet{HoffSep} for the array normal model.
\end{longlist}

\subsection{Accommodating missing data}\label{secmiss}
Mortality information is limited for many undeveloped countries that do
not have reliable death registration data. Thus, it is not uncommon to
be missing a country's death rates for specific ages or at all ages in
a given year. Both the maximum likelihood and Bayesian estimation
procedures can be modified to accommodate missing data, however, such
modifications are often computationally expensive.

In the maximum likelihood estimation, expectation--maximization
algorithms are often employed to obtain parameter estimates in the
presence of missing data. The proposed coordinate ascent algorithm with
an additional step that computes the expectation of the log probability
of the data under the SFA model given the current values of the
parameters and observed data would correspond to an
expectation/conditional maximization (ECM) algorithm [\citet
{MengRubin}]. \citet{GAllen} discuss such an algorithm in detail
for the
matrix normal ($K=2$), when additional penalties are placed on the
covariance matrices, and find that such an algorithm is not
computationally feasible for high-dimensional data due to the
complicated expectations required. An analogous algorithm for array
data and the SFA model would likely suffer from the same burdens.
\citet{GAllen} further propose an approximation of the ECM
procedure to
obtain estimates of the missing values that involves the following
three steps: initialize the missing values, compute maximum likelihood
estimates of the parameters, and use an iterative procedure to compute
the expectation of the missing values conditional on the observed data
and parameters. While an analogous approximation could be developed for
the SFA model, the procedure for the matrix case lacks theoretical
guarantees and was also shown to require complete iteration of all three steps to obtain estimates
that sufficiently match those from the ECM.

Accommodating missing data in a Bayesian framework is straightforward
and provides predictive distributions for the missing values. The
proposed Metropolis--Hastings algorithm can easily be adapted by
including additional steps that sample portions of the missing data
from their full conditional distributions. Although the full
conditional distribution of all missing data elements conditional on
the parameters and observed data can be expressed as a multivariate
normal, calculating the parameters for this distribution is often
computationally expensive due to the large matrices involved in
computing the distribution's covariance matrix. However, using results
from \citet{HoffSep}, the conditional distribution of a slice of an
array (where one mode index is fixed) can be written as an array normal
distribution. The missing data within the slice conditional on the
observed data in the slice follows a multivariate normal distribution,
which can be sampled from to update the missing values. Calculating the
conditional distribution of the missing elements in a slice of the
array via this two-step conditioning procedure (once for the slice and
once for the missing data within the slice) circumvents computation
with unnecessarily large matrices. \citet{GAllen} used a similar procedure
to obtain expected values of missing elements in a matrix normal model
in their ECM approximation. Section~\ref{sechmd} illustrates the use of
a Metropolis--Hastings algorithm that has been modified to accommodate
and provide predictions for missing mortality data.

\subsection{Testing for the mode ranks} \label{sectest}

It is often difficult to choose the number of factors for a single mode
factor model. This problem is only more pronounced in the array case
where the rank $k_{i}$ must be specified for each mode. As in single
mode factor analysis [\citet{MKB}], a likelihood ratio test can be
constructed to test between nested SFA models with ranks $(k_{1},\ldots,k_{K})$ and $(k_{1}^{*},\ldots,k_{K}^{*})$, where $k_{i} \le
k_{i}^{*}$ for all $i$. However, due to the large number of possible
combinations of ranks, choosing the ranks using these likelihood ratio
tests is challenging. In the Bayesian framework, alternative approaches
to specifying the factor rank in single mode factor analysis are to
estimate it along with the model parameters using MCMC estimation
methods such as reversible jump [\citet{LopesWest2004}] and path
sampling [\citet{LeeSong2002}], or specify an infinite number of factors
[\citet{BhattacharyaDunson}]. While it is possible to extend these
methods to the array case and SFA model, they would greatly increase
the computational complexity of estimation. Maximum likelihood
parameter estimates via the coordinate ascent algorithm can be obtained
in minutes even for a large array such as the mortality data, while the
MCMC Bayesian estimation procedure can take hours to run depending on
the size of the array and complexity of the mean model. Therefore, here
we propose an alternative mode-by-mode rank selection procedure based
on the maximum likelihood parameter estimates that suggests when the
rank specified for a given mode is sufficient for capturing the
dependence within that mode.

As in Section~\ref{secmode}, let $\widetilde{Y}$ denote a $K$-way array
that has been standardized by all mode covariance matrices. To
determine whether the dependence in mode $i$ is captured by a proposed
$(k_{1},\ldots,k_{K})$ SFA model, we can compute $\widetilde{Y}$ using
the SFA mode covariance matrix estimates as in (\ref{SFA}) and test
whether the covariance matrix of the rows of $\widetilde{Y}_{(i)}$
equals the identity.
The likelihood ratio test statistic for this test~is
%
\begin{equation}
t = \frac{m}{m_{i}} \bigl[ \operatorname{tr}(\widehat{V}) - \log
|\widehat{V}|
- m_{i} \bigr], \label{LRstat}
\end{equation}
where\vspace*{1pt} $\widehat{V} = \frac{m_{i}}{m}\widetilde{Y}_{(i)}\widetilde
{Y}_{(i)}^{T}$, and
has an asymptotic $\chi^{2}_{m_{i}(m_{i}+1)/2}$ distribution under the
null hypothesis of an identity row covariance matrix. Note that
rejecting this test suggests that a more complex covariance structure
is needed for the $i$th mode. This motivates the following rank
selection procedure for the entire array:
\begin{longlist}[3.]
\item[0.]
Consider an SFA model with all $k_{i}=0$. Obtain estimates of the
covariance parameters $D_{i}$ using the maximum likelihood procedure in
Section~\ref{secMLE} and compute $\widetilde{Y}$ using the estimates.
\item[1.]
For each mode $i$, define $R_{i} = \operatorname{Cov}[\operatorname
{vec}(\widetilde{Y}_{(i)})]$ and test $H_{0}\dvtx R_{i} = \mathrm
{I}_{m/m_{i}} \otimes\mathrm{I}_{m_{i}} $ vs $H_{1}\dvtx R_{i} =
\mathrm
{I}_{m/m_{i}} \otimes V $, where $V$ is an unstructured $m_{i} \times
m_{i}$ covariance matrix, using a likelihood ratio test with test
statistic given by (\ref{LRstat}).
\item[2.]
If the test for mode $i$ rejects and
\[
\cases{ \delta(m_{i},k_{i}+1) > 0, &\quad increase the rank
$k_{i}$ by one,
\vspace*{2pt}\cr
\delta(m_{i},k_{i}+1) \le0, &
\quad set the rank equal to $m_{i}$.}
\]

If the test for mode $i$ does not reject, fix $k_{i}$ at its current
value and perform no further tests on the mode. Obtain maximum
likelihood estimates $\{\widehat{\bolds{\Sigma}},\widehat{\bolds
{\Lambda}},\widehat{\mathbf{D}}\}$ for an SFA model with
the new ranks $(k_{1},\ldots,k_{K})$ and compute $\widetilde{Y}$ using
these new estimates.
\item[3.]
Repeat steps 1--2 until each mode has failed to reject a test.
\end{longlist}
The suggested ranks $(k_{1},\ldots,k_{K})$ are those that result at
the end
of this procedure. In step~2 $\delta(m,k) = [ (m-k)^{2} - (m+k)
]/2$ represents the reduction in the number of parameters when
using a $k$-factor analytic covariance matrix instead of an $m \times
m$ unstructured covariance matrix. When $\delta(m,k) \le0$, a factor
analytic covariance structure no longer provides a reduction in the
number of covariance parameters and an unstructured covariance matrix
should be specified. Note that if a nonzero mean model was specified,
its parameters $\beta$ would be simultaneously estimated with the
covariance matrices at each iteration of the procedure.

The maximum number of SFA models that could be considered using this
procedure is bounded by the largest value of $k_{l}$ such that $\delta
(m_{l},k_{l})>0$, where $l$ denotes the array mode with the largest
dimension $m_{l}$. To control the type~I error rate of all mode tests
to be $\alpha$ for an iteration of steps 1~and~2, the level of each
mode test can be set to $\alpha^{r}$, where $r$ is the number of modes
being tested (i.e., the number that have rejected every test thus far).
An example of this procedure is described in Section~\ref{sechmdcov}
for the mortality data.

\section{Application to Human Mortality Database death rates} \label{sechmd}
In this section we analyze death rates from the Human Mortality
Database (HMD) using an SFA model, compare our model to other
covariance models, and obtain predictions for over four hundred missing
death rates. We focus on death rates for 5-year time periods for
populations corresponding to combinations of sex, age and country of
residence. Specifically, we consider death rates from 1960 to 2005 for
40 countries, both sexes and twenty-three age groups, $\{
0$, 1--4, 5--9, 10--14$,\ldots,105+\}$. These data are represented in a 4-way
array $Y=\{y_{\mathit{ctsa}}\}$ of dimension $(40 \times9 \times2 \times23)$,
where $y_{\mathit{ctsa}}$ is the log death rate for country $c$, time period
$t$, sex $s$ and age group~$a$. We will refer to a set of age-specific
death rates for a combination of country, time period and sex as a
mortality curve.

We begin this section by introducing a flexible piecewise polynomial
mean model and show the residuals from this mean model exhibit
dependence within each mode: age, time period, country, and sex. Using
the likelihood ratio testing procedure presented in Section~\ref
{sectest}, we select ranks for an SFA model. The resulting SFA model is
compared to models with simpler covariance structures using
out-of-sample cross-validation and is used to impute multiple years of
missing death rates for Chile and Taiwan.

\subsection{Mean model selection}

As discussed in the \hyperref[secintro]{Introduction}, existing methods for analyzing
mortality data model the death rates for different countries, sexes
and/or time periods separately. Such an approach can be inefficient due
to the strong similarities between mortality rates within the same
country, time period and sex. For this reason, we propose a new joint
mean model for the HMD data that exploits these relationships between
mortality rates that share levels of one or more of these factors.

Figure~\ref{MortCurves} shows mortality curves defined by the
twenty-three age-specific death rates for the United States and Sweden
in four time periods. The large spikes at age zero represent infant
mortality, and the humps around age twenty, which are especially
evident in males, are attributed to teenage and young adult accident
mortality. The overall shapes of the mortality curves for each sex are
similar across countries and time periods, however, Sweden has
considerably lower mortality levels during childhood and young
adulthood compared to the United States. This suggests that a mean
model for the data should allow for different curves across countries
and time periods, yet still take advantage of the similarity between
death rates within the same country, age group or sex.

Drawing from the mortality literature and viewing mortality rates as
function of age, we propose the following piecewise polynomial (PP)
mean model:
%
\begin{eqnarray}
\label{mean} \E[y_{\mathit{ctsa}}] = \cases{ \phi^{0}, &\quad $a =0$,
\vspace*{2pt}\cr
\phi^{1} + a \phi^{11} + a^{2}\phi^{12}, &\quad $1 \le a < 20$,
\vspace*{2pt}\cr
\phi^{2} + a\phi^{21}+ a^{2}\phi^{22} + a^{3}\phi^{23},&\quad $20\le a,$}
\nonumber\\[-25pt]\\[10pt]
\eqntext{\phi^{i} = \alpha_{\c}^{i} + \beta_{t}^{i} + \gamma_{s}^{i}.}
\end{eqnarray}
This model distinguishes between the infant, childhood and adult stages
of mortality by fitting each with a separate polynomial, whose
coefficients are composed of additive effects for country, time period
and sex. The constant term at age zero is necessary to model the steep
decline from infant mortality to child mortality that is not well
represented by a low degree polynomial.

One of the most commonly used models in demography for age-specific
mortality measures is the Heligman--Pollard (HP) model [\citet
{HP}]. This
model also uses eight parameters to parameterize a mortality curve,
however, it is typically used to model each mortality curve
individually and is nonlinear and nonconvex in the parameters, making
estimation extremely difficult [\citet{Hartmann1987}, \citet
{Congdon1993}]. When the HP model is fit separately to the 684 HMD
mortality curves for the 38 countries missing no death rates using OLS,
it requires over 5400 parameters and under the assumption of
independent, homoscedastic errors has a Bayesian Information Criterion
(BIC) value of $-$17,288. However, when the PP model is fit jointly to
the same data using OLS, it contains 376 parameters and has a BIC of $-$52,436.
Due to the relative parsimony of the PP model, its superior fit in
terms of BIC, and its straightforward estimation as a linear model, it
was selected as the mean model.

\subsection{Excess dependence and SFA rank selection}\label{sechmdcov}

The piecewise polynomial model in (\ref{mean}) is extremely flexible.
To investigate its fit to the HMD mortality rates, we focused on a
subset of the original data that contains no missing observations,
specifically the $(38 \times9 \times2 \times23)$ array that does not
contain death rates for Chile or Taiwan. The OLS fit explains 99.5\% of
the variation in the mortality rates (coefficient of determination,
$R^{2} = 0.995$). However, there is interest in whether excess
correlation exists in the residuals since modeling it can improve both
predictions of missing values and the efficiency of parameter
estimates. OLS estimates of the parameters in (\ref{mean}) are
equivalent to maximum likelihood estimates assuming independent normal
errors. To evaluate this latter assumption, we computed the empirical
correlation matrix for each mode based on the mean model residuals.

As mentioned in the \hyperref[secintro]{Introduction}, the distributions of these
correlations have substantially more large positive values than would
be expected under the assumption of independent errors. For example,
speaking specifically to the temporal dependence, the average
correlation between adjacent time periods, those one time period apart
and those two periods apart is 0.79, 0.54 and 0.26, respectively. The
first two principal components of each correlation matrix are shown in
Figure~\ref{pcs}. The horseshoe pattern in the time period principal
components and the clustering of countries within the same region
suggest temporal and geographic trends in the data are not captured by
the mean [\citet{Horseshoe}]. This indicates that even though the mean
model contains several country-specific and time period-specific
parameters, similarities between the mortality curves of certain
countries and time periods is not being accounted for. The mean model
already contains over 370 parameters and it would likely be nontrivial
to modify it to capture all of the dependence seen in the residuals.
For this reason, we consider incorporating a covariance structure to
model this excess dependence. An array normal separable covariance
structure could be specified, however, it would add over one thousand
parameters to the model. Therefore, we instead consider an SFA model
for the data with the PP mean with the belief that the residual
dependence within some modes may be well approximated by a low rank
factor analytic structured covariance matrix.

As outlined in Section~\ref{sectest}, suggestions for the SFA ranks can
be obtained from a repeated likelihood ratio testing procedure. For the
mortality data, we consider $(k_{\c},k_{t},k_{s},k_{a})$ SFA models
where the ranks correspond to the country, time period, sex and age
covariance matrices, respectively. The standardized residual array
$\widetilde{Y}$ for a $(k_{\c},k_{t},k_{s},k_{a})$ SFA model is defined
as $\operatorname{vec}(\widetilde{Y}) = (\operatorname{vec}(Y) -
\operatorname{vec}(\widehat{M}))(\widehat{\Sigma}_{a}^{-1/2}
\otimes
\widehat{\Sigma}_{s}^{-1/2} \otimes\widehat{\Sigma}_{t}^{-1/2}
\otimes
\widehat{\Sigma}{}^{-1/2}_{\c})$, where $\widehat{M}$ represents the PP
mean model estimate and $\widehat{\Sigma}_{i}$ is the SFA mode $i$
covariance matrix estimate from the maximum likelihood estimation
procedure (modified to estimate the mean model and covariance
parameters simultaneously). The results from the iterative testing
procedure are shown in Table~\ref{LRT}. The first step in this process
is to consider a $(0,0,0,0)$ SFA model where all covariance matrices are
diagonal. The likelihood ratio test statistics for this model are shown
in the first row of Table~\ref{LRT} and the corresponding 0.05 level
critical values are shown in the last row. Since the test for each mode
rejects the null hypothesis of independent, variance one errors, the
rank of each mode is increased by one in the subsequent model, except
for that for the sex mode. A rank one factor analytic structure for a
$(2 \times2)$ covariance matrix has more parameters than an
unstructured covariance matrix, so the sex covariance matrix is
unstructured in the next model. A box around a test statistic in the
table indicates the mode failed to reject the test for the first time.
Recall that when a mode's test does not reject, the rank for that mode
is fixed and not increased in later models. The table shows where the
sex, time period, country and age ranks become fixed at two, four, nine
and ten, respectively. Observe that after a mode's rank is fixed, the
test statistic for that mode stays below the critical value in all
subsequent models. Although the mode tests are not independent of the
covariance structures fit in the other modes, this consistency supports
the suggested ranks.

%
\begin{table}
\tabcolsep=0pt
\caption{Iterative testing procedure for the SFA ranks. Each row represents
an SFA model and each entry is the likelihood ratio test statistic
based on (\protect\ref{LRstat}). The 0.05 level critical value for each
test is given in the last row. A box around a statistic indicates that
the mode does not reject the test for the first time and the rank is
fixed in subsequent models}\label{LRT}
\begin{tabular*}{\tablewidth}{@{\extracolsep{\fill}}@{}ld{5.0}d{5.0}d{3.0}d{5.0}@{}}
\hline
\multirow{2}{56pt}{\mbox{}\vspace*{-5.5pt}\break\textbf{SFA ranks}\hfill{\break} $\bolds{(k_{c},k_{t},k_{s},k_{a})}$} & \multicolumn{4}{c@{}}{\textbf{Likelihood ratio test statistic}}\\[-6pt]
& \multicolumn{4}{c@{}}{\hrulefill}\\
& \multicolumn{1}{c}{\textbf{Country}} & \multicolumn{1}{c}{\textbf{Time period}} & \multicolumn{1}{c}{\textbf{Sex}} &  \multicolumn{1}{c@{}}{\textbf{Age}}\\
\hline
$(0,0,0,0)$ & 21\mbox{,}852 & 14\mbox{,}482 & 702 & 27\mbox{,}883\\
$(1,1,2,1)$ & 9526 & 5853 &\multicolumn{1}{c}{\phantom{0,}\fbox{0}} & 14\mbox{,}451 \\
$(2,2,2,2)$ & 4425 &1722 &0 &6374 \\
$(3,3,2,3)$ & 2776 &716 & 0 &3762 \\
$(4,4,2,4)$ & 1946 &\multicolumn{1}{c}{\phantom{00,}\fbox{17}} &0 &2422 \\
$(5,4,2,5)$ &1556 &14 & 0 &1833 \\
$(6,4,2,6)$ & 1287 &10 & 0 & 1340\\
$(7,4,2,7)$ & 1040 & 8 & 0 & 967\\
$(8,4,2,8)$ & 892 &5& 0 & 540\\
$(9,4,2,9)$ & \multicolumn{1}{c}{\phantom{0,}\fbox{762}} & 8 & 0& 363\\
$\bolds{(9,4,2,10)}$ &737 & 8 & 0 & \multicolumn{1}{c@{}}{\phantom{0,}\fbox{257}}
\\[6pt]
$\chi^{2}_{0.95}$ critical value & 805 & 62 & 8 & 316\\
\hline
\end{tabular*}
\end{table}

\subsection{Out-of-sample cross-validation}
We evaluate the SFA model by comparing its out-of-sample predictive
performance with two simpler covariance models that share the same PP
mean model. The three covariance models considered are the following:
\begin{longlist}[M3:]
\item[M1:] Independent and identically distributed (i.i.d.) model.

\item[M2:] Time covariance model.

\item[M3:] SFA model $(9,4,2,10)$.
\end{longlist}
M1 corresponds to the conventional ordinary least squares (OLS)
approach where all errors are assumed independent and identically
distributed with a common variance. In general, country mortality rates
are relatively stable over time, so if the observed mortality for a
given country, year and age deviates from the mean model in one year,
it is likely the observations deviate in the same direction in
neighboring years. Thus, a natural first step to incorporating a
covariance model is to consider an unstructured covariance matrix for
time as in M2.

Fifty cross-validations were performed by removing a random 25\% of the
array, estimating each of the three covariance models with the PP mean
model on the remaining data, and computing the mean squared error (MSE)
between the observed values and the predicted values for the withheld
entries. The predicted values for M1 are those from the OLS PP mean
estimate. For M2 and M3, the predictions are the posterior mean
estimates of the missing values from the Bayesian estimation procedure
described in Section~\ref{secMCMC}, modified to accommodate missing data.

%
\begin{table}[b]
\tabcolsep=0pt
\tablewidth=300pt
\caption{Average and standard deviation of the mean squared errors from
50 out-of-sample cross-validation~experiments}\label{MSE}
\begin{tabular*}{\tablewidth}{@{\extracolsep{\fill}}@{}lccc@{}}
\hline
& \textbf{M1 (i.i.d.)} & \textbf{M2 (time covariance)} & \textbf{M3 (SFA)}\\
\hline
Average MSE                & 0.02996 & 0.00729 & 0.00385\\
Standard deviation of MSEs & 0.00084 & 0.00049 & 0.00034 \\
\hline
\end{tabular*}
\end{table}

A prior distribution for the parameters in the PP model is needed to
perform simultaneous Bayesian estimation for the mean and covariance
parameters. The prior on the vector of PP coefficients is a mean-zero
normal distribution with covariance matrix $m(X^{T}X)^{-1}$, where $X$
is the design matrix for the PP model for $\operatorname{vec}(Y)$ and
$m=\prod_{i=1}^{K} m_{i}$. This is a relatively uninformative prior, as
it is over 30 times more diffuse than the corresponding
unit-information prior [\citet{KassWass}]. The hyperparameters were
specified as described in Section~\ref{secMCMC} where the mean estimate
$\widehat{M}$ used in $\hat{\psi}$ is the OLS estimate of the PP
model. Since M2 has no modes with factor analytic structure, the prior
on the time covariance matrix is
\[
\Sigma_{t}^{-1} \sim\operatorname{Wishart} \biggl(
n_{t}=m_{t}+2, \frac
{m\hat{\psi}}{m_{t}} \mathrm{I}_{m_{t}} \biggr).
\]
This specification is necessary to preserve the property that
$\E[\operatorname{tr}(\operatorname{Cov}[\operatorname{vec}(Y)])] =
\hat{\psi}$ under the prior.

The results from the 50 cross-validations are shown in Table~\ref{MSE}.
The MSE for the SFA model was less than that of the time covariance
model for each of the 50 cross-validations, and
the MSE for the time covariance model
was always less than that of the i.i.d. model.
In terms of average MSE, both the time covariance model and the SFA
model significantly improve upon the i.i.d. model, and the SFA model
outperforms the time covariance model by nearly a factor of two.
This is evidence that even with the extremely flexible PP mean model,
the SFA covariance structure still improves model fit, as it is able to
estimate the similarity between mortality rates across countries, time
periods, age groups and sexes, and use this information in its predictions.

\subsection{Prediction of missing data}\label{secpred}
The imputation of missing death rates is an important application of
modeling mortality data, as information is often incomplete for
countries lacking accurate death registration data. We now consider the
original $(40 \times9 \times2 \times23)$ array of mortality rates
with observations for Chile and Taiwan. Seven time periods of mortality
information are missing for Chile (1960--1995) and two time periods for
Taiwan (1960--1970), combining for a total of 414 missing entries in the
array. This larger data array contains only two additional countries,
so the SFA ranks $(9,4,2,10)$ selected for the reduced data are used
again here. Predictions for the missing death rates were based on
samples from the Metropolis--Hastings procedure, for which the effective
sample sizes
for the Monte Carlo estimates of
all missing values were greater than 500.

%
\begin{figure}[b]

\includegraphics{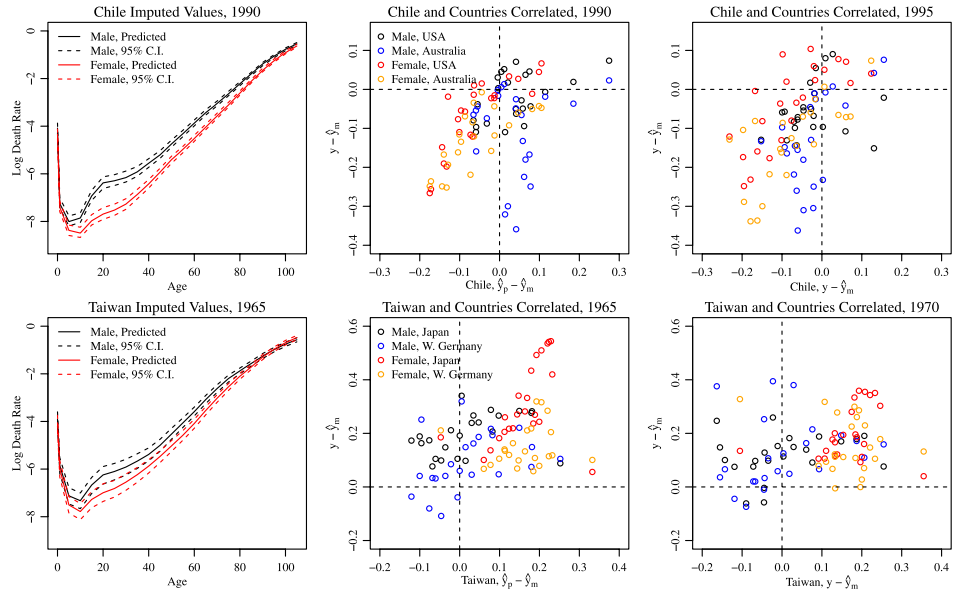}

\caption{The first column of plots shows the predicted values and
corresponding 95\% prediction intervals for the missing death rates for
Chile and Taiwan. The middle column shows the difference between the
posterior mean predicted value and the piecewise polynomial mean
function fitted value, $\hat{y}_{p} - \hat{y}_{m}$, for Chile and
Taiwan, along with empirical mean model residuals, $y - \hat{y}_{m}$,
for countries that are highly correlated with them in the posterior
mean country covariance matrix. The last column contains empirical
residuals for the following time period when Chile and Taiwan mortality
is observed.}\label{pred}
\end{figure}

In the left column of Figure~\ref{pred}, posterior mean predicted death
rates and 95\% prediction intervals are shown for Chile in 1990 and
Taiwan in 1965. To visualize the impact of the SFA covariance model on
the predicted death rates, we investigate the difference between the
SFA predicted values and the fitted values based on the PP mean model.
The SFA predictions, $\hat{y}_{p}$, are conditional on the observed
mortality rates for all other countries and time periods, while the
mean model fitted values, $\hat{y}_{m}$, are based only on the estimate
of the PP mean model. These differences, $\hat{y}_{p} - \hat{y}_{m}$,
are called ``predictive residuals'' since they are based on predicted
values instead of observed values and illustrate the changes in the
predicted values by using the SFA covariance model compared to only
using the mean model. The empirical residuals based on the PP mean
model, $y - \hat{y}_{m}$, were computed for
the United States and Australia,
the two countries most highly correlated with Chile (estimated
correlations around 0.40).
These residuals were also computed for Japan and West Germany, the two
countries most highly correlated with Taiwan (estimated correlations of
around 0.13).
The middle column of Figure~\ref{pred} shows the predictive residuals
for Chile and Taiwan and the empirical residuals for these select
countries. The last column contains the empirical residuals in 1995 and
1970 when mortality information is available for all countries. Observe
that the plots in the middle column and last column are similar,
demonstrating an overall positive association for both sexes and all
country pairs. This demonstrates how the model uses the relationship
between the empirical residuals of Chile and other countries to predict
Chile's deviations from the mean model in years when Chile data is
missing. The ability to draw information across multiple country, year
and sex residuals to impute missing values is a critical strength of
the SFA model that is not shared by other mortality models or simpler
covariance structures.

The empirical residuals for Chile shown in the last column may not show
as strong of an association with the United States and Australia as one
would expect from a posterior mean correlation estimate of 0.4.
However, recall that the estimate of the country correlations is based
on all time periods, sexes and ages. Although we show adjacent time
periods in this plot, the correlation between the country residuals in
the period adjacent to the missing time period and the correlations in
time periods furthest away are weighted equally in the estimate of the
country correlation, and hence weighted equally in the imputation of
the missing data. This property is a consequence of the separability of
the SFA covariance matrix. A more complicated nonseparable covariance
model would be required for the correlations between countries, ages
and sexes to be differentially weighted in the imputation based on the
proximity of the observed data to the missing data.

\section{Discussion}\label{secdis}
In this article we introduced the separable factor analysis model for
array-valued data. Unlike the array normal model where all mode
covariance matrices are unstructured, SFA parameterizes mode covariance
matrices by those with factor analytic structure. Using covariance
matrices with reduced structure decreases the number of parameters in
the model considerably and allows mode covariance matrices to be
estimated using maximum likelihood methods for any array dimension.
Including a covariance structure in a model for multiway data can
drastically improve mean model parameter estimation and missing data
predictions in situations where dependence exists within modes that is
not captured by the mean model. In an out-of-sample cross-validation
study with a large set of mortality data, the SFA model was shown to
have superior fit compared to models with simpler covariance
structures, even in the presence of an extremely flexible mean model.
The SFA model was also shown to estimate which countries have similar
deviations from the mean model and was able to use this information in
its predictions of multiple years of missing death rates.

%
\begin{figure}[b]

\includegraphics{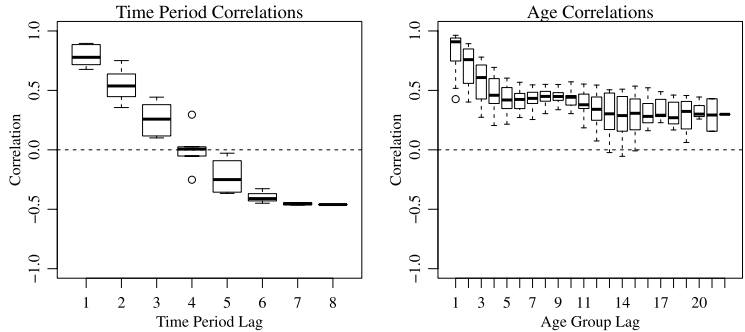}

\caption{Sample residual correlations between time periods and age
groups from the OLS fit of the model in (\protect\ref{mean}) grouped
by lag.}\label{laggedcors}
\end{figure}

We propose reducing the number of covariance parameters in the array
normal model by modeling mode covariance matrices with factor analytic
structure, however, other simplified covariance structures are
possible. For example, the Bayesian graphical lasso [\citet{Wang2012}]
and covariance matrices derived from Gaussian graphical models
[\citet{WangWest}, \citet{DobraLenkoskiRodriguez}] estimate
or assume
conditional independencies between pairs of indices (reflected by zeros
in the precision matrix) and are commonly used to represent covariance
among index sets which have a natural spatial structure. Similarly, for
temporal data, a covariance matrix derived from an autoregressive model
is commonly used. Nevertheless, in many cases there may not be a clear
choice of a reduced structured covariance matrix, and in these cases
specifically we propose the factor analytic structure as an agnostic
approach to covariance matrix parameter reduction. It was suggested
that an autoregressive covariance structure may be appropriate for the
time and/or age mode covariance matrices in the mortality data
application in Section~\ref{sechmd}. Figure~\ref{laggedcors} shows
boxplots of the residual correlations discussed in the \hyperref[secintro]{Introduction},
grouped by time period lag and age group lag. If an autoregressive
model of order-1 were appropriate for either of these modes, we would
expect the correlations to monotonically decrease toward zero with lag.
However, the negative correlations exhibited by the time periods and
asymptoting behavior of the age correlations cannot be captured by such
structure. This illustrates that even in instances when traditional
covariance structures may seem appropriate, they may not be given the
mean model, and it may be preferable to take a more agnostic approach
to modeling and assume a factor analytic structure.

A trivial extension of the SFA model would be to relax the separability
assumption for groups of modes of the array. For example, in the
mortality data, if we believed the residual correlation between sexes
and across time periods was not separable, the four-way array could be
unfolded into a three-way array whose dimensions are age, country and
time period/sex. An SFA model could then be specified for the resulting
three-way array.
Relaxing the separability assumption between some modes is likely to
improve model fit for specific data sets when the assumption of
separability is not appropriate, however, this also increases the
potential number of covariance parameters in the corresponding array
normal model. Therefore, in order to gain a sufficient reduction in the
number of covariance parameters, a small factor model rank is likely to
be necessary for combined modes. Investigation of the empirical
residual correlations may help suggest when relaxing the separable
assumption is warranted.

\begin{appendix}
\section*{Appendix: Sampling \texorpdfstring{$\Lambda$}{Lambda} and $D$}\label{appMH}
Let $\Lambda_{i}^{*}$ be the proposed value of $\Lambda$ that results
from 2(a--b). The acceptance probability for this proposal is
\begin{eqnarray*}
\alpha\bigl(\Lambda_{i}^{*},\Lambda_{i} \bigr)
&=& \frac{p(\Lambda_{i}^{*}|Y,\Lambda
_{-i},D,\Sigma)p(\Lambda_{i}|\Lambda_{i}^{*},D,\Sigma,\Lambda
_{-i},Y)}{p(\Lambda_{i}|Y,\Lambda_{-i},D,\Sigma)p(\Lambda
_{i}^{*}|\Lambda_{i},D,\Sigma,\Lambda_{-i},Y)}
\\
&=& \frac{p(Y|\Lambda_{i}^{*},\Lambda_{-i},D,\Sigma)p(\Lambda
_{i}^{*}|D_{i})p(\Lambda_{i}|\Lambda_{i}^{*},D,\Sigma,\Lambda
_{-i},Y)}{p(Y|\Lambda_{i},\Lambda_{-i},D,\Sigma)p(\Lambda
_{i}|D_{i})p(\Lambda_{i}^{*}|\Lambda_{i},D,\Sigma,\Lambda_{-i},Y)}.
\end{eqnarray*}
The proposal probability can be written
\begin{eqnarray*}
&& p \bigl(\Lambda^{*}_{i} | \Lambda_{i},D,\Sigma,
\Lambda_{-i},Y \bigr)
\\
&&\qquad = \int p \bigl(\Lambda_{i}^{*},
Z^{i}| \Lambda_{i}, D,\Sigma,\Lambda_{-i},Y \bigr)
\,dZ^{i}
\\
&&\qquad = \int p \bigl(\Lambda_{i}^{*} | Z^{i},D,\Sigma,
\Lambda_{-i},Y \bigr) p \bigl(Z^{i}| \Lambda_{i},
D, \Sigma,\Lambda_{-i},Y \bigr) \,dZ^{i}
\\
&&\qquad = p \bigl(\Lambda_{i}^{*}| D,\Sigma,\Lambda_{-i},Y
\bigr)
\\
&&\quad\qquad{}\times  \int\frac{
p(Z^{i}|\Lambda_{i}^{*}, D,\Sigma,\Lambda_{-i},Y)}{ p(Z^{i}|
D,\Sigma,\Lambda_{-i},Y)} p \bigl(Z^{i}| \Lambda_{i}, D,
\Sigma,\Lambda_{-i},Y \bigr) \,dZ^{i}
\\
&&\qquad = \frac{p(Y|\Lambda_{i}^{*}, D,\Sigma,\Lambda_{-i})p(\Lambda_{i}^{*},
D,\Sigma,\Lambda_{-i})}{p(D,\Sigma,\Lambda_{-i},Y) } \cdot c \bigl
(\Lambda_{i},\Lambda^{*}_{i}|D,
\Sigma,\Lambda_{-i},Y \bigr)
\\
&&\qquad = \frac{p(Y|\Lambda_{i}^{*}, D,\Sigma,\Lambda_{-i})p(\Lambda
_{i}^{*}|D_{i})p(D)p(\Sigma)p(\Lambda_{-i}|D_{-i})}{p(D,\Sigma,\Lambda
_{-i},Y) }
\\
&&\quad\qquad{}\times  c \bigl(\Lambda_{i},\Lambda^{*}_{i}|D,
\Sigma,\Lambda_{-i},Y \bigr),
\end{eqnarray*}
where $c(\Lambda_{i},\Lambda^{*}_{i}|D,\Sigma,\Lambda_{-i},Y)$
represents the integral, which is symmetric in $\Lambda_{i}$ and
$\Lambda_{i}^{*}$. Plugging the last expression into the acceptance
probability, we obtain $ \alpha(\Lambda_{i}^{*},\Lambda_{i}) = 1$.
Analogous logic can be used to show the acceptance probability for a
proposed $D_{i}$ from $2(c-d)$ is also one.
\end{appendix}

\section*{Acknowledgments}
The authors would like to thank the Editor, Associate Editor and
referees for their advice on this manuscript.



\printaddresses

\end{document}